\documentclass[aps,preprint,amsmath,amssymb,floatfix]{revtex4}
\usepackage[dvips]{epsfig}
\usepackage{graphicx}
\usepackage{subfigure}
\usepackage{dcolumn}
\usepackage{bm}
\usepackage{color}
\usepackage[utf8]{inputenc}

\newcommand{\black}{\textcolor{black}}

\begin{document}
\hyphenation{va-ni-sh-ing}

\begin{center}
{\large\bf Thermopower enhancement by encapsulating\\ cerium in clathrate cages}
\\[0.5cm]

A.\ Prokofiev$^{1\ast}$, A.\ Sidorenko$^1$, \black{K.\ Hradil$^2$,} M.\ Ikeda$^1$,
R.\ Svagera$^1$, M.\ Waas$^1$, H.\ Winkler$^1$,\\ K.\ Neumaier$^3$,
and S.\ Paschen${^{1\ast\ast}}$\\

$^1$Institute of Solid State Physics, Vienna University of Technology,\\ Wiedner Hauptstr.\ 8-10, 1040 Vienna, Austria\\

\black{$^2$X-Ray Center, Vienna University of Technology,\\  Getreidemarkt 9,
1060 Vienna, Austria}\\

$^3$Walther-Mei{\ss}ner-Institute for Low Temperature Research,\\ Walther-Mei{\ss}ner-Str.\ 8,  85748 Garching, Germany
\end{center}

\vspace{0.5cm}

\noindent {\bf The increasing worldwide energy consumption calls for the design
of more efficient energy systems. Thermoelectrics could be used to convert waste
heat back to useful electric energy if only more efficient materials were
available. The ideal thermoelectric material combines high electrical
conductivity and thermopower with low thermal conductivity. In this regard, the
intermetallic type-I clathrates show promise with their exceedingly low lattice
thermal conductivities \cite{Tob08.1}. Here we report the successful
incorporation of cerium as guest atom into the clathrate crystal structure. In
many simpler intermetallic compounds, this rare earth element is known to lead,
via the Kondo interaction, to strong correlation phenomena including the
ocurrence of giant thermopowers at low temperatures \cite{Pas06.1r}. Indeed, we
observe a 50\% enhancement of the thermopower compared to a rare earth-free
reference material. Importantly, this enhancement occurs at high temperatures
and we suggest that a `rattling' enhanced Kondo interaction \cite{Hot07.1}
underlies this effect.}

\vspace{3cm}

\noindent $^{\ast}$ e-mail: prokofiev@ifp.tuwien.ac.at\\
\noindent $^{\ast\ast}$ e-mail: paschen@ifp.tuwien.ac.at

\newpage

\noindent Type-I clathrates are guest--host systems with the general composition
G$_8$H$_{46}$. The guest atoms G are situated in polyhedral face-sharing and
space-filling cages formed by the tetrahedrally bonded host atoms H
(Fig.\,\ref{fig1}a). Exceptionally low lattice thermal conductivities are an
intrinsic property of these materials as they are observed even in pure single
crystalline specimens \cite{Tob08.1}. They have been attributed to reduced
phonon group velocities resulting from the interaction of acoustic and rattling
modes \cite{Chr08.1,Euc12.1}. The charge carriers are essentially unaffected by
these lattice anomalies. Thus type-I clathrates appear to be a realization of
the `phonon glass--electron crystal' concept \cite{Sla95.1}. With this in mind,
many groups worldwide have synthesized and investigated a large number of type-I
clathrates with various compositions over the past decades. As a result a
significant increase has been achieved in the thermoelectric figure-of-merit
$ZT=S^2\sigma/\kappa\cdot T$, where $T$ is the absolute temperature, $S$ the
thermopower, $\sigma$ the electrical conductivity, and $\kappa$ the thermal
conductivity. The highest $ZT$ values reported are 1.63 at 1100\,K for $n$-type
Ba$_8$Ga$_{16}$Ge$_{30}$ (ref.\,\onlinecite{Sar06.1}) and 1.1 at 900\,K for
$p$-type Ba$_8$Ga$_{16}$Ge$_{30}$ (ref.\,\onlinecite{Ann02.1}).

Record high power factors $S^2\sigma$ have, however, been found in a very
different class of materials---in strongly correlated rare earth or transition
metal compounds \cite{Row02.1,Sal94.1,Jie12.1,Pas06.1r}. The giant thermopower
values observed in these systems can be traced back to an enhanced quasiparticle
density of states (DOS) near the Fermi level which results from the Kondo
interaction of the local 4$f$ states with the conduction electrons
\cite{Zla05.1,Col07.1}.

Thus, the incorporation of adequate rare earth elements into type-I clathrates
appears as promising \cite{Pas06.1r} yet to date unachieved
\cite{Kaw00.1,Tan08.1,Kov11.1} route to superior thermoelectric
materials. In particular, it has not been possible to incorporate any sizable
amount of Ce---the rare earth element with strongest tendency to hybridize with
conduction electrons---into a clathrate phase \cite{Kaw00.1}.
Here we report on the first successful synthesis of a Ce-containing clathrate
and on the finding that its thermopower is sizeably enhanced with respect to a
rare earth-free reference material.

In our search for a synthesis route to a Ce-containing clathrate, we were guided
by the Zintl-Klemm concept \cite{Sch85.1} which predicts that polar
intermetallic compounds are stable if all atoms reach filled electron shells. In
type-I clathrates one assumes that the host atoms are covalently bound to each
other while the guest atoms are ionically bound to the host. For the usual case
of electropositive guest atoms this is realized by a transfer of the guest
atom's valence electrons to the host, which thus becomes a polyanion. An example
is the clathrate (M$^{+2}$)$_8$(III$^{-1}$)$_{16}$(IV$^{0}$)$_{30}$, where the
2-valent metal atoms M are the guest atoms and the group III and group IV
elements form the framework. The superscripts represent the formal charges of
the atoms: the valence electrons of M are donated to the 3-valent elements III
to establish the fourth framework bond. With all valence electrons used up an
electrical insulator is expected. Real clathrates are semiconductors or poor
metals; this is attributed to slight deviations from the above oversimplified
situation.

A particularly broad composition range around the ideal Zintl composition was
found in the transition metal clathrate Ba$_8$Au$_x$Si$_{46-x}$ (BAS), that
forms in a wide Au concentration range $2.2\leq x \leq6$
(ref.\,\onlinecite{Ayd11.1}) including the ideal Zintl composition
(Ba$^{+2}$)$_8$(Au$^{-3}$)$_{5.\dot 3}$(Si$^{0}$)$_{40.\dot 7}$ for monovalent
Au. This points to an exceptional robustness of this phase, which is why we have
chosen it as starting material for our study.

The substitution of two-valent Ba by three-valent Ce must be accompanied by an
increase of the acceptor capacitance of the framework and thus by an enhanced Au
content. At $x = 6$, the Ce content for charge balance corresponds to the
clathrate (Ba$^{+2})_6$(Ce$^{+3})_2$(Au$^{-3})_6$(Si$^0)_{40}$ with two Ce atoms
per formula unit. This ideal Zintl composition was the starting point for our
synthesis (Methods), which finally yielded single crystalline clathrate
specimens of the composition Ba$_{6.91 \pm 0.17}$Ce$_{1.06 \pm 0.12}$Au$_{5.56
\pm 0.25}$Si$_{40.47 \pm 0.43}$ (Ce-BAS, see Fig.\,\ref{fig1} and Methods). The
presence of the sizable amount of about 1 Ce atom per formula unit in the
clathrate phase is unambiguously proven by our EDX and WDX analysis, together
with the fact that no foreign phase is present (Fig.\,\ref{fig1}b, \ref{fig1}c),
even on the nanometer scale (Fig.\,\ref{fig1}d). It is also in line with the
sizable reduction of the lattice parameter\black{, evidenced by the refinement of powder X-ray diffraction (XRD) data sets (Methods),} compared to BAS samples
with similar Au content. We also prepared a phase pure La-containing clathrate
(Ba$_{6.99 \pm 0.17}$La$_{1.23 \pm 0.12}$Au$_{5.91 \pm 0.25}$Si$_{39.87 \pm
0.43}$, La-BAS, Methods) as $4f$-free reference compound. \black{Structural refinements using single crystal XRD data sets (Supplementary Information) reveal that both Ce and La
occupy the $2a$ site within the smaller cage (Fig.\,\ref{fig1}a).}

Having firmly established the presence of Ce (and La) in our single phased
clathrate samples by analytical and structural investigations we now turn to
their physical properties. The electrical resistivity $\rho(T)$ of
La-BAS decreases with decreasing temperature, behaviour typical for metals. By
contrast, $\rho$ of Ce-BAS {\em in}creases upon cooling (Fig.~\ref{fig2}a). This
`semiconductor-like' characteristic is unexpected, as discussed below.

The linear response Hall coefficient $R_{\rm H}(T)$ is negative for Ce-BAS and
positive for La-BAS, in agreement with the Zintl count based on the measured
compositions. The absolute value of $R_{\rm H}$ is distinctly smaller for Ce-BAS
than for La-BAS (Fig.\,\ref{fig2}b), suggesting that Ce-BAS has a higher charge
carrier concentration and should thus be more metallic than La-BAS. The
temperature dependence of $|R_{\rm H}|(T)$ is similar for both compounds. It is
typical of heavily doped semiconductors. In particular, no hint for an anomalous
contribution to the Hall effect \cite{Fer87.1} of Ce-BAS can be discerned. Thus,
the enhanced resistivity of Ce-BAS is not a charge carrier concentration effect
but is instead due to a reduced Hall mobility $\mu_{\rm H}=|R_{\rm H}|/\rho$
(Fig.\,\ref{fig2}b, inset).

A comparison with published mobility data for a series of polycrystalline BAS
samples \cite{Ayd11.1} reveals that $\mu_{\rm H}$ of our La-BAS single crystal
is significantly enhanced with respect to the $p$-type BAS sample closest in
carrier concentration. This is most naturally attributed to reduced scattering
in a single crystal due to the absence of grain boundaries, an effect that was
also seen in Ba-Ni-Ge clathrates \cite{Ngu10.1}. By contrast, the mobility of
our single crystalline Ce-BAS sample is significantly lower than \black{that of}
the polycrystalline $n$-type BAS sample of similar carrier concentration. Thus,
an additional scattering process must be evoked to explain the low mobility of
Ce-BAS.

The most obvious mechanism for a metal containing Ce ions is Kondo scattering.
In ordered heavy-fermion metals, incoherent Kondo scattering above the single
ion Kondo temperature $T_{\rm K}$ leads to a resistivity contribution
proportional to $-\ln T$. Due to the temperature dependent charge carrier
concentration of Ce-BAS this dependence can only be expected to hold
approximately, in agreement \black{with} our results (see Fig.\,\ref{fig2}a).
Below $T_{\rm K}$, ordered heavy-fermion metals show a pronounced decrease of
the resistivity, which is due to the onset of coherence in a Kondo lattice. The
absence of this decrease in Ce-BAS is explained as follows. The multiplicity of
the $2a$ site, \black{at which the Ce atoms are situated, is 2. As Ce-BAS contains
only 1.06 Ce ions per formula unit it does not occupy this site fully.} Thus,
there is a sizable amout of Kondo holes \cite{Nak02.1} on the lattice which will
inhibit the formation of a coherent state. Magnetic susceptibility and specific
heat measurements presented below support the importance of Kondo phenomena in
Ce-BAS.

\black{Next, we discuss the thermoelectric properties. The thermopower $S(T)$ is
the quantity that is at the very heart of the phenomenon of thermoelectricity.
It} is negative (positive) for Ce-BAS (La-BAS, Fig.~\ref{fig2}c), in agreement
with the Hall effect results presented above. Extrema of $-180$\,$\mu$V/K
(300\,$\mu$V/K) are reached at 480\,K (375\,K) for Ce-BAS (La-BAS). These values
exceed those of all previously studied BAS clathrates \cite{Can12.1,Zei12.1}.
\black{In the next paragraph we will show that, even though La-BAS reaches higher
$S$ values, it is in fact Ce-BAS that shows an anomalous thermopower
enhancement.} The highest power factors $S^2\sigma$ reached in the temperature
range 2-600\,K are approximately 11\,$\mu$WK$^{-2}$cm$^{-1}$ for La-BAS at 350~K
and 6.0\,$\mu$WK$^{-2}$cm$^{-1}$ for Ce-BAS at 600~K, with a tendency to further
enhancement at higher temperatures (Fig.~\ref{fig2}d). \black{The thermal
conductivities $\kappa(T)$ of Ce-BAS and La-BAS (Fig.~\ref{fig2}e, inset) are
similar to the ones found in the BAS series \cite{Can12.1}. Since $\kappa(T)$ is
dominated by the lattice contribution $\kappa_{\rm l}$ this shows that the
rattling modes of Ce-BAS and La-BAS are just as effective in producing a low
lattice thermal conductivity as in BAS. For $n$-type Ce-BAS, the maximum $ZT$
value of about 0.15 at 480\,K (0.19 at 600\,K if a constant $\kappa_{\rm l}$
above 480\,K is assumed) is 30\% (100\% at 600\,K) higher than that of the best
$n$-type BAS material at the same temperature. For $p$-type La-BAS, $ZT$ is 0.2
at 400\,K, which is 35\% higher than that of the best $p$-type BAS sample at the
same temperature.}

In simple metals and degenerate semiconductors the thermopower depends on the
charge carrier concentration as $S \propto n^{-2/3}$. In Fig.\,\ref{fig3} we
show such a plot for the room temperature thermopower of Ce-BAS and La-BAS,
together with data for $p$- and $n$-type BAS samples from the literature
\cite{Ayd11.1}. While La-BAS nicely fits into the series, Ce-BAS has a strongly
enhanced thermopower: $|S|$ is by as much as 50\% enhanced with respect to the value expected from its carrier
concentration. May this again be attributed to the Kondo interaction? In order
to answer this question we have performed thermodynamic measurements at low
temperatures, that are likely to reveal more insight into the Kondo physics of
this system.

The magnetic susceptibility $\chi(T)$ reveals that Ce-BAS is paramagnetic, while
La-BAS is diamagnetic (Fig.\,\ref{fig4}a), with no sign of magnetic ordering
down to the lowest accessed temperatures. The paramagnetic behaviour of Ce-BAS
can be attributed to the presence of Ce$^{+3}$ ions. They have the electron
configuration [Xe]$4f^1$ and thus possess one localized, well shielded electron
that carries a magnetic moment. The six-fold degenerate spin-orbit ground state
$^2F_{5/2}$ is generally split by crystal electric fields. In cubic point
symmetry (the situation relevant for the $2a$ site), a $\Gamma_7$ doublet and a
$\Gamma_8$ quartet result. The temperature dependence of an ensemble of free
(non-interacting) moments will be determined by the thermal population of the
different energy levels. Since the splitting between the $\Gamma_7$ and the
$\Gamma_8$ state is typically of the order of 50 to $100\,$K, the crystal field
ground state will govern the lowest temperature properties.

Between about 1 and 6\,K, $\chi(T)$ of Ce-BAS shows Curie-Weiss type behaviour,
with an effective moment of $1.48\,\mu_{\rm B}$ per Ce ion and a paramagnetic
Weiss temperature close to zero (Fig.\,\ref{fig4}a, inset). The latter is
expected due to the large distance between nearest neighbouring Ce atoms of
about 0.9\,nm. Below 1\,K, the magnetic susceptibility deviates from this
Curie-Weiss law. Its tendency to saturate to a constant value can be attributed
to the Kondo interaction.

This is further corroborated by the specific heat $C(T)$ of Ce-BAS. It shows a
pronounced anomaly in zero magnetic field (Fig.\,\ref{fig4}b) that can be
attributed to the splitting of a Kramers doublet by the Kondo interaction. Both
the peak temperature and the peak value increase with applied fields, again
behaviour typical for Kondo systems. After subtracting the phonon specific heat
of the reference compound La-BAS, the magnetic entropy of the anomaly can be
estimated. An entropy of $0.4 R \ln{2}$ per mole-Ce, where $R$ is the universal
gas constant, is released up to 0.5 K. Twice this temperature is generally
considered a good estimate of the Kondo temperature \cite{Mel82.1,Geg08.1}.
Taken all information together, Ce-BAS behaves as an incoherent Kondo system
with a low Kondo temperature of $T_{\rm K} \approx 1$\,K at low temperatures. 

This raises an important question: How can this low Kondo temperature lead to an
enhancement of the thermopower even above room temperature? We argue in the
following that a second, much higher Kondo scale emerges in this system as a
consequence of spin-phonon coupling. 

First we discuss the evidence for rattling in both Ce-BAS and La-BAS.
Einstein-like contributions to the specific heat can be nicely revealed by
$C_p/T^3$ vs $\ln T$ plots, where they appear as bell-shaped anomalies on top of
a background contribution due to a Debye solid plus charge carrier/spin
contributions at low temperatures \cite{Jun83.1}. By the introduction of Ce and
La, the peak observed for BAS is only slightly reduced in amplitude and slightly
shifted to higher temperatures (Fig.~\ref{fig4}c). \black{This can be attributed
to a slightly reduced amount of free space \cite{Sue07.1} for the Ba ions in the
small cages, the diameter of which have shrunk by 0.81\% (0.89\%)  by the
introduction of the smaller ion Ce (La) (Supplementary Information).
Surprisingly, the atomic displacement parameter at the $2a$ site shows a
relative enhancement in Ce-BAS (La-BAS) compared to values for BAS
(Supplementary Information). This indicates that the rattling amplitude of the
Ce (La) ions in the small cages is enhanced compared to that of Ba at the same
site in BAS.} Thus, rattling is a characteristic of Ce-BAS (La-BAS) just as it
is for rare earth-free clathrates. In the following we argue that this has
profound implications on the Kondo interaction.

One of the simplest models that treats the interplay between localized electrons
and local optical phonon modes is the Anderson-Holstein model \cite{Hew02.1}. It
has been shown that, in the regime of strong electron correlation, the Kondo
energy scale can be strongly enhanced by the electron-phonon interaction
\cite{Hot07.1}. Rattling in clathrates is a thermally activated process. Thus,
local phonon modes are populated only at elevated temperatures. This may lead to
a strong renormalization of the Kondo energy scale with increasing temperature.
We argue below that the temperature dependence of the magnetic susceptibility
supports this picture.

The paramagnetic contribution to the magnetic susceptibility of Ce-BAS (Methods)
is shown as $\chi T$ vs $T$ in Fig.\,\ref{fig4}d. A pure Curie law corresponds
to constant behaviour in this plot. We see two regions with weak temperature
dependence, one up to about 6\,K (plateau 1) and a second one between about 120
and 400\,K (plateau 2). Qualitatively, the transition between plateau 1 and 2 is
most naturally attributed to the thermal population of the upper crystal
electric field (CEF) level. However, the susceptibility reached at plateau 2 is
sizably reduced with respect to the local moment limit (free Ce$^{+3}$ ions
would give $\chi T \approx 0.81$\,emu\,K/mol). Also the increase of the
susceptibility above plateau 2 does not fit into a simple CEF level
scheme. Instead, the suppression of the susceptibility from the local moment
limit, that has already set in at the highest temperature of 700\,K accessed in
our experiment, can be attributed to a Kondo temperature of several hundred
Kelvin.

Thus, the following picture emerges. The rattling modes of the Ce ions
encapsulated in the cages of the clathrate
Ba$_{6.91}$Ce$_{1.06}$Au$_{5.56}$Si$_{40.47}$ couple to the conduction
electrons on the framework, thereby effectively reducing the Coulomb repulsion
$U$, and enhancing the system's Kondo temperature by orders of magnitude. This
leads to both the early suppression of the magnetic susceptibility from the
high-temperature local moment limit and to the 50\% enhancement of the
thermopower (\black{and 100\% enhancement of $ZT$ at 600\,K if $\kappa_{\rm
l}$(600\,K$) = \kappa_{\rm l}$(480\,K) is assumed)} over the value of the rare earth-free
reference material. We expect these results to open up a new avenue for research
in thermoelectrics. Also, they are likely to stimulate further research into
related phenomena, such as magnetically robust \cite{San05.3,Cos12.1} and
electric dipolar Kondo behaviour \cite{Hot12.1}, or thermopower enhancements in
the negative-$U$ Anderson model \cite{And11.1}.

\vspace{1cm}

{\small
{\noindent\large\bf Methods}\\
{\noindent\bf Synthesis, structural and analytical characterization.} Polycrystalline
starting material was synthesized by melting of high purity Ba, Ce, Au and Si in
a cold copper boat using high-frequency heating. The resulting samples were
polyphased, with the clathrate-I phase being the majority phase. Long term
annealing did not change the phase relations. The amount of Ce detected by
energy-dispersive X-ray spectroscopy (EDX) in the clathrate phase was, with
about 1.2 at.\% (0.65 Ce per formula unit), smaller than in the starting
material.

Phase pure Ce-containing clathrates (Ce-BAS) were finally obtained by
off-stoichiometric crystal growth by the floating-zone melting technique using
optical heating in a four-mirror furnace (Crystal Corporation). The obtained
clathrate sample of about 5$\times$4$\times$3 mm$^3$ consisted of a few single
crystals and was phase pure according to powder X-ray diffraction (XRD, Siemens
D5000 diffractometer, Fig.\,\ref{fig1}b) and scanning electron microscopy (SEM,
Philips XL30 ESEM, Fig.\,\ref{fig1}c). High resolution transmission electron
microscopy (HRTEM, FEI TECNAI F20) proved the absence of nano-inclusions
(Fig.\,\ref{fig1}d). The La-containing Ba-Au-Si clathrate (La-BAS) was
prepared with a similar technique.

\black{Rietveld refinements of the powder XRD patterns of powdered
single crystalline samples yield the lattice parameters $a=1.0395(2)$\,nm and
$a=1.0392(2)$\,nm for Ce-BAS and La-BAS, respectively. These are sizeably
smaller than published values for Ba$_8$Au$_x$Si$_{40-x}$ clathrates with
similar Au content \cite{Her99.1,Ayd11.1,Zei12.1}. Since the ionic radii of Ce
and La are smaller than the one of Ba this supports the incorporation of the
rare earth atoms in the clathrate phase.} Also energy and wave length dispersive
X-ray analysis  (EDX: EDAX New XL-30 135-10 UTW+ detector, WDX: Microspec
WDX-600), with a single phase Ba$_8$Au$_5$Si$_{41}$ sample as standard, confirm
the presence of Ce and La in the crystals. The average crystal compositions
measured by EDX are Ba$_{6.91 \pm 0.17}$Ce$_{1.06 \pm 0.12}$Au$_{5.56 \pm
0.25}$Si$_{40.47 \pm 0.43}$ and Ba$_{6.99 \pm 0.17}$La$_{1.23 \pm
0.12}$Au$_{5.91 \pm 0.25}$Si$_{39.87 \pm 0.43}$. \black{Single crystal XRD investigations are reported in Supplementary Information.}

{\noindent\bf Physical property measurements.} The magnetic properties below room
temperature were measured in a superconducting quantum interference device
(SQUID) magnetometer (Cryogenics S700X) after cooling down the samples in zero
applied field. Above room temperature, the susceptibility was measured by the
vibrating sample magnetometer (VSM) option of a Physical Property Measurement
System (PPMS, Quantum Design). The specific heat below 2\,K was measured with a
relaxation-type method using a home-built set up in a $^3$He/$^4$He dilution
refrigerator. Above 2\,K, it was measured by the same technique in a PPMS.
Electrical resistivity and Hall effect measurements were done by standard
4-point and 6-point techniques, with alternating dc current in a PPMS. Below
room temperature, thermopower was determined by a two-heater method in a
home-built cryostat. Above 300~K, thermopower was measured simultaneously
with the electrical resistivity using a standard 4-point steady-state dc
technique (ZEM 3, Ulvac-Riko). \black{The thermal conductivity was determined from thermal diffusivity measurements using a standard laser-flash experiment
(Anter, Flashline 3000FT - S2) and measured specific heat data.}

{\noindent\bf Diamagnetic susceptibility contributions.} To correct the magnetic
susceptibility of Ce-BAS for diamagnetic contributions we use the $4f$-free
reference material La-BAS, which is expected to have very similar diamagnetic
contributions. Its susceptibility has the form
\begin{equation}
\chi = \chi_{\rm{d,cs}} + \chi_{\rm{d,ring}} + \chi_{\rm{p,imp}} +
\frac{2}{3}\chi_{\rm{p,Pauli}}
\end{equation}
where $\chi_{\rm{d,cs}}$ and $\chi_{\rm{d,ring}}$ are diamagnetic contributions
of closed shells
and due to ring currents \cite{Her99.1},
respectively. $\chi_{\rm{p,imp}}$ is a paramagnetic contribution due to a small
amount of magnetic impurities; it is relevant only at the lowest temperatures.
$\chi_{\rm{p,Pauli}}$ is the Pauli paramagnetic contribution of the conduction
electrons. The factor 2/3 accounts for Landau diamagnetism. We determine
$\chi_{\rm{p,Pauli}}$ and $\chi_{\rm{p,imp}}$ of La-BAS from Hall effect
measurements and a fit to a Curie-Weiss law at low temperature, respectively,
and then subtract both these terms from the total measured $\chi(T)$ data of
La-BAS. The result is identified as the total diamagnetic contribution
$\chi_{\rm{d}}$ of both La-BAS and Ce-BAS.


\vspace{0.8cm}


\vspace{0.8cm}

{\noindent\large\bf Acknowledgements}\\
We acknowledge financial support from the Austrian Science Fund (projects
P19458-N16\black{, TRP 176-N22, and I623-N16}) and the European Research Council (Advanced Grant QuantumPuzzle, no.\ 227378)\black{, and help from T.\ Pippinger and R.\ Miletich-Pawliczek (Vienna University) during single crystal XRD measurements on La-BAS.}

\vspace{0.8cm}

{\noindent\large\bf Author contributions}\\
A.P.\ and S.P.\ designed the research. A.P.\ synthesized the material, A.S.,
M.I., R.S., M.W., H.W., K.N.\ \black{and K.H.\ }performed the measurements. A.P., A.S.,
H.W., \black{K.H.\ }and S.P.\ analysed the data. A.P., A.S.\ and S.P.\ prepared the
manuscript.

\vspace{0.8cm}

{\noindent\noindent\large\bf Additional Information}\\
The authors declare that they have no competing financial interests.
Reprints and permission information is available
online at http://www.nature.com/reprints. Correspondence and requests for
materials should be addressed to A.P.\ and S.P.


\newpage
\begin{figure}[h!]
\caption{\label{fig1}{\bf $|$ Structure and quality characterization of the clathrate Ce-BAS.}
{\bf a}, Schematic representation of the crystal structure of type-I clathrates.
The two different cages are sketched. The guest atoms (G, red) sit at the $2a$
($6d$) site with cubic (tetragonal) point symmetry in the smaller (larger) cage
made up of host atoms (H, blue). {\bf b}, The powder X-ray diffraction pattern
(top) can be fully indexed with the type-I clathrate structure (bottom). {\bf
c}, Scanning electron microscopy images at two different magnifications reveal
that no micro-inclusions are present. {\bf d}, The representative transmission
electron microscopy image rules out inhomogeneities on the nanometer scale.}
\end{figure}

\begin{figure}[h!]
\caption{\label{fig2}{\bf $|$ Transport properties of Ce-BAS and La-BAS.}
Temperature dependencies of {\bf a}, the electrical resistivity {\bf b}, the
absolute value of the Hall coefficient {\bf b}, {\bf inset}, the Hall mobility,
{\bf c}, the thermopower, \black{ {\bf d}, the power factor, {\bf e}, the
dimensionless thermoelectric figure of merit, and {\bf e}, {\bf inset}, the
thermal conductivity (symbols, left axis), the electronic part $\kappa_{\rm e}$ estimated via the Wiedemann-Franz law from $\rho(T)$ (full lines, left axis), and the lattice thermal conductivity $\kappa_{\rm l}=\kappa - \kappa_{\rm e}$ (dashed lines, right axis). The
black lines are guides to the eye. They were used to calculate $\mu_{\rm H}$ and $ZT$.}}
\end{figure}

\begin{figure}[h!]
\caption{\label{fig3}{\bf $|$ Thermopower comparison with rare earth-free
reference material BAS.} Room temperature thermopower of Ce-BAS and La-BAS,
together with published values for a series of BAS samples with different Au
content \cite{Ayd11.1} as function of the charge carrier concentration, plotted
as $n^{-2/3}$. The lines are linear fits to the data for BAS.}
\end{figure}

\begin{figure}[h!]
\caption{\label{fig4}{\bf $|$ Thermodynamic properties of the investigated
clathrates.} {\bf a}, Magnetic susceptibility of Ce-BAS (top) and La-BAS
(bottom) vs temperature. {\bf a}, {\bf inset}, Inverse magnetic susceptibility
of Ce-BAS at low temperatures. {\bf b}, Low-temperature specific heat of Ce-BAS
in different magnetic fields. {\bf c}, Specific heat of Ce-BAS, La-BAS and BAS
plotted as $C/T^3$ vs temperature. {\bf d}, Temperature dependence of the
paramagnetic contribution to the magnetic susceptibility of Ce-BAS, plotted as
$\chi\,T$ vs temperature.}
\end{figure}


\newpage

\begin{figure}[t!]
\centerline{\includegraphics[width=180mm]{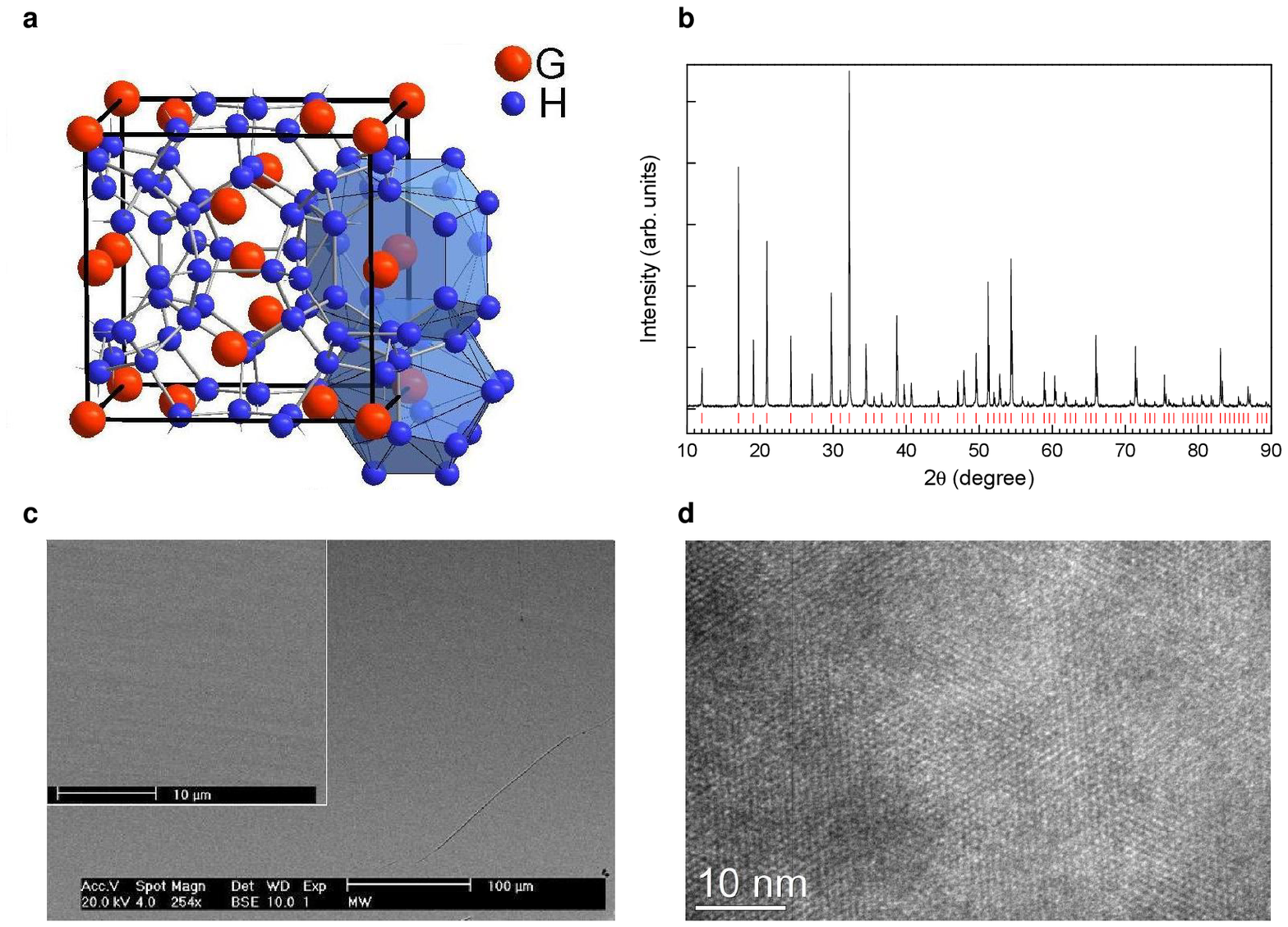}}
\vspace{1cm}

{\Large Figure 1}
\end{figure}

\newpage

\begin{figure}[t!]
\centerline{\includegraphics[width=180mm]{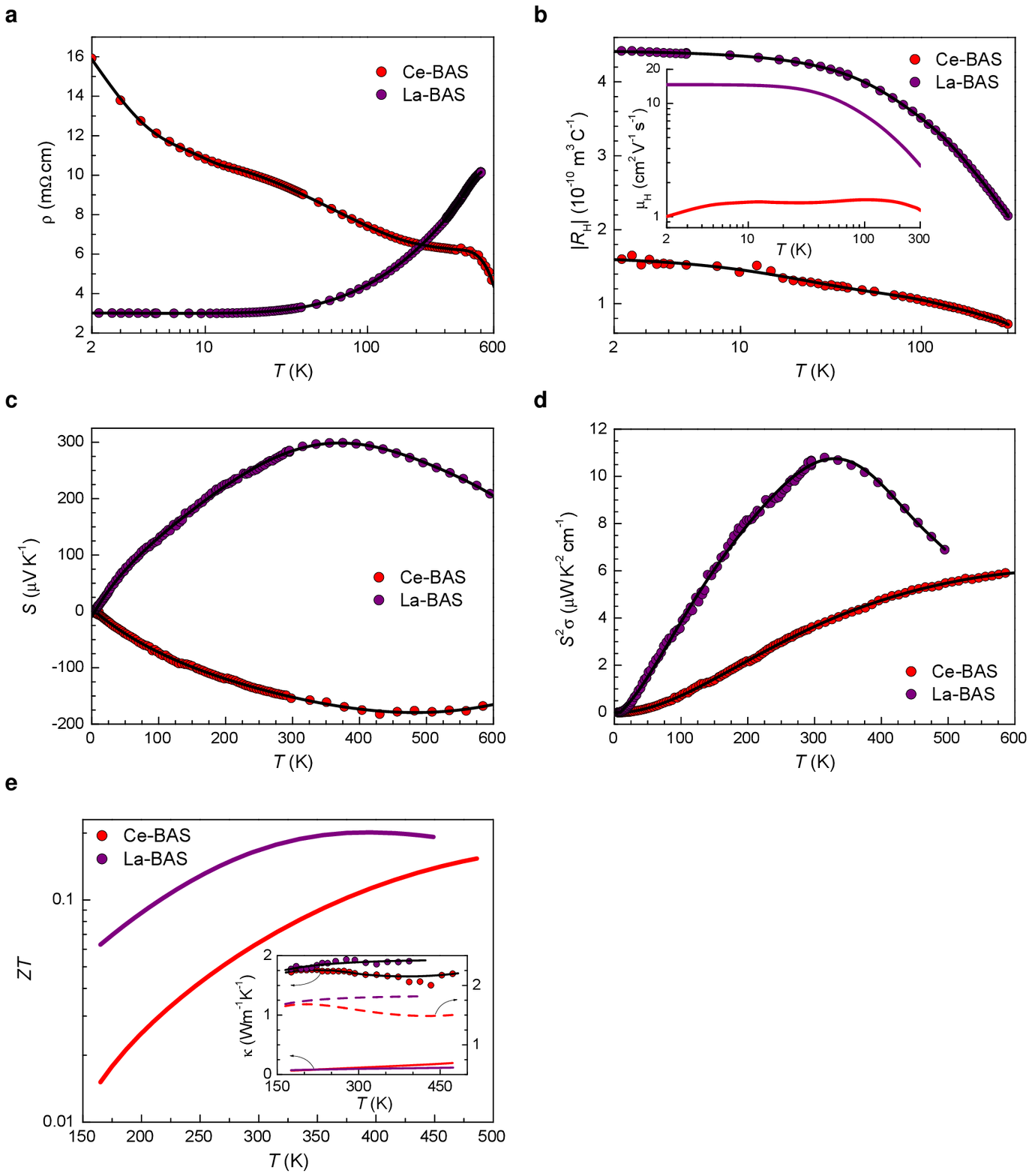}}
\vspace{1cm}

{\Large Figure 2}
\end{figure}

\newpage

\begin{figure}[t!]
\centerline{\includegraphics[width=140mm]{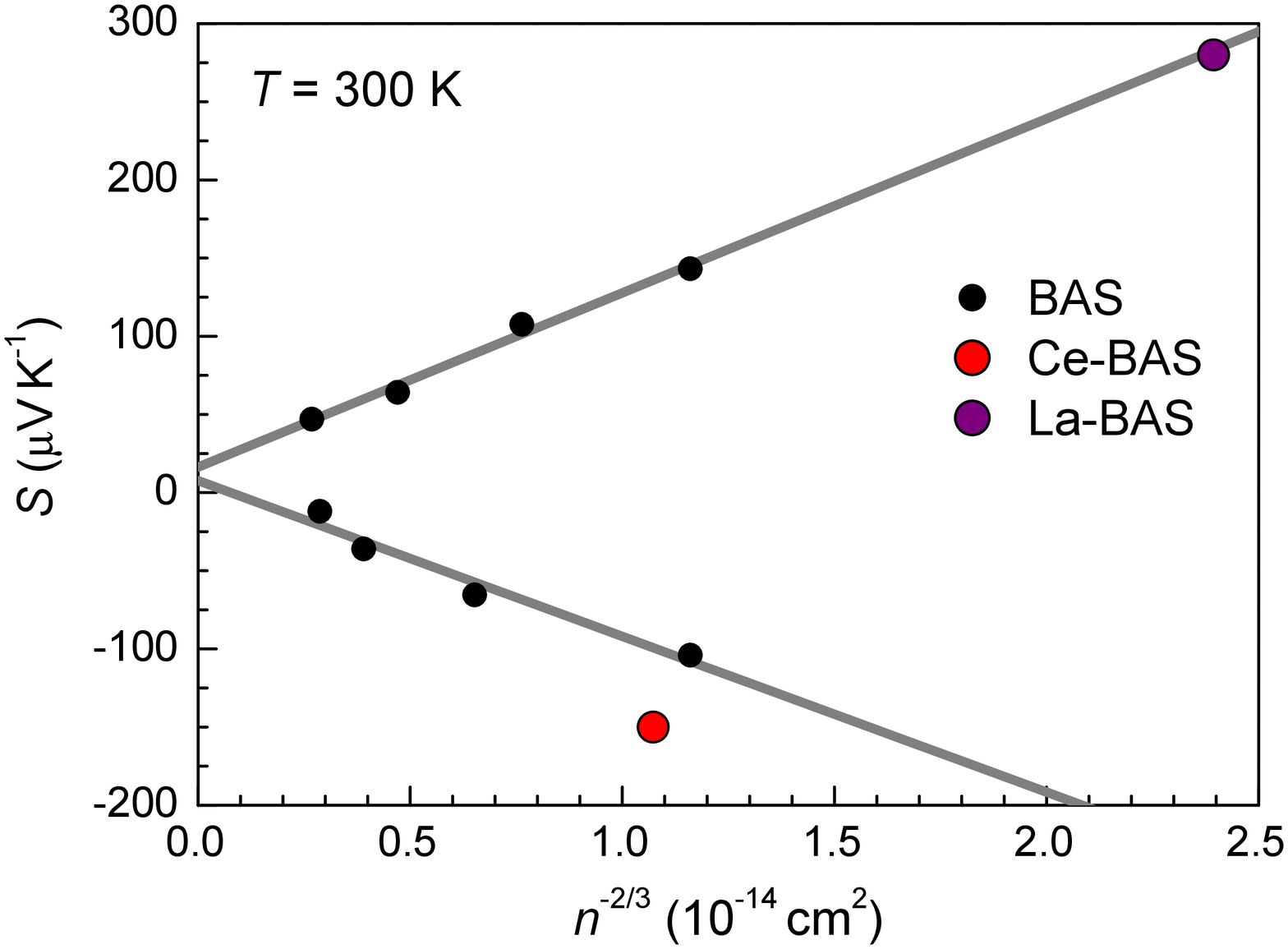}}
\vspace{1cm}

{\Large Figure 3}
\end{figure}

\begin{figure}[t!]
\centerline{\includegraphics[width=180mm]{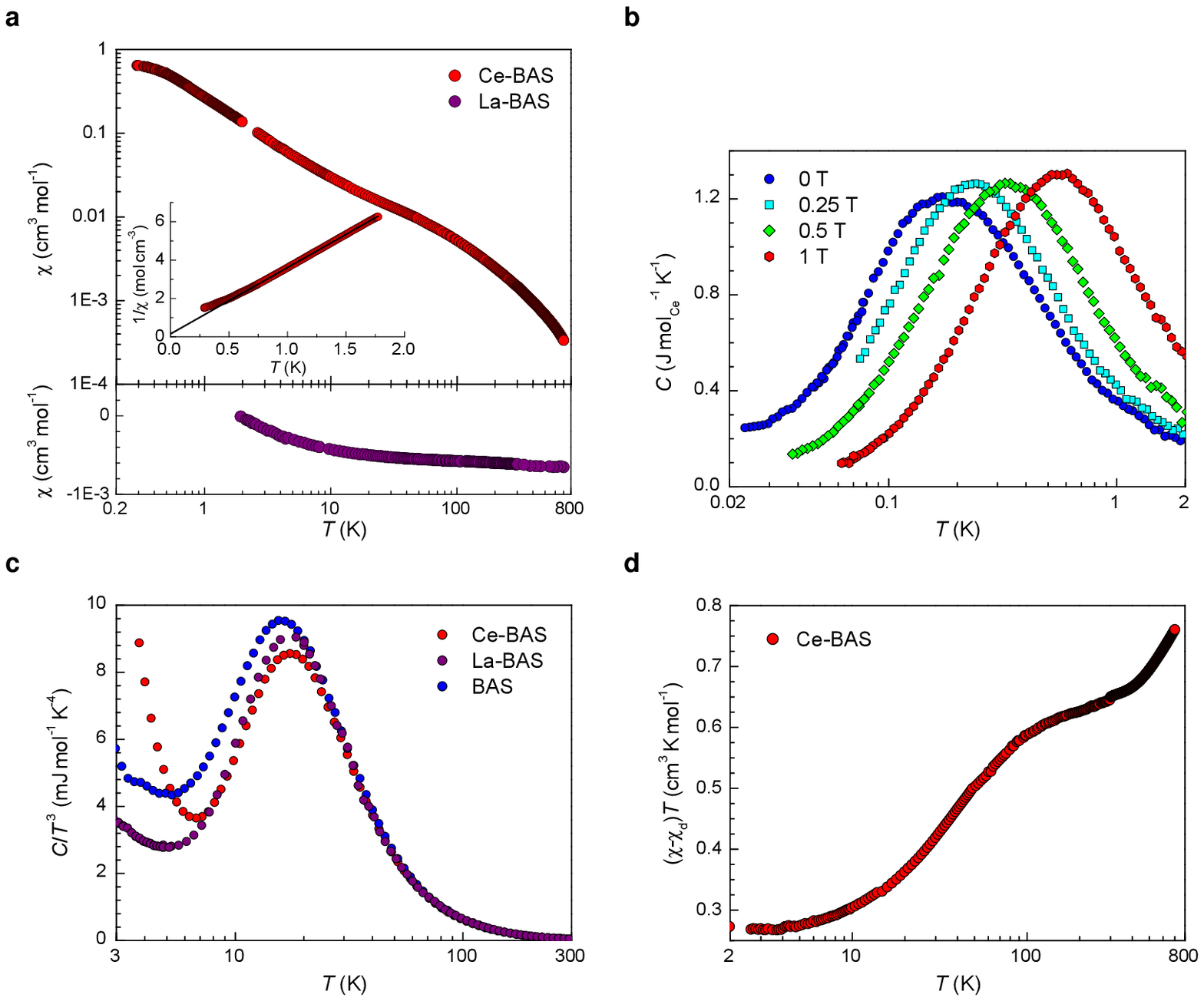}}
\vspace{1cm}

{\Large Figure 4}
\vspace{8cm}
\end{figure}

\end{document}